\pdfoutput=1
\documentclass[10pt,conference]{IEEEtran}

\usepackage{xspace}
\usepackage{booktabs}   
\usepackage{subcaption} 
\usepackage{xcolor}
\usepackage{textcomp}
\usepackage{pgfpages}
\usepackage{pgfplots}
\usepackage{tikz}
\usepackage{enumerate}
\usepackage{algorithmicx}
\usepackage{algorithm}
\usepackage[noend]{algpseudocode}
\usepackage{multirow}
\usepackage{amsmath}
\usepackage{listings}
\usepackage{colortbl}
\usepackage{url}
\usepackage[font=bf]{caption}
\usepackage{flushend}

\usepackage{hyperref}
\hypersetup{breaklinks,hidelinks,pdftitle={Harvey: A Greybox Fuzzer for Smart Contracts},pdfauthor={Valentin Wuestholz,
Maria Christakis},pdfkeywords={greybox fuzzing, dynamic program
analysis}}
\definecolor{gray}{rgb}{0.5, 0.5, 0.5}
\definecolor{light-gray}{gray}{0.77}
\definecolor{BrickRed}{rgb}{0.8, 0.25, 0.33}
\definecolor{Black}{rgb}{0.0, 0.0, 0.0}
\definecolor{DarkBlue}{rgb}{0.0, 0.0, 0.55}
\definecolor{Crimson}{rgb}{0.86, 0.08, 0.24}
\definecolor{SlateGrey}{rgb}{0.44, 0.5, 0.56}
\definecolor{lightorange}{HTML}{FFB74D}
\definecolor{blue}{rgb}{0.0, 0.0, 1.0}
\definecolor{magenta}{rgb}{0.79, 0.08, 0.48}

\makeatletter
\newenvironment{btHighlight}[1][]
{\begingroup\tikzset{bt@Highlight@par/.style={#1}}\begin{lrbox}{\@tempboxa}}
{\end{lrbox}\bt@HL@box[bt@Highlight@par]{\@tempboxa}\endgroup}

\newcommand\btHL[1][]{%
  \begin{btHighlight}[#1]\bgroup\aftergroup\bt@HL@endenv%
}
\def\bt@HL@endenv{%
  \end{btHighlight}%
  \egroup
}
\newcommand{\bt@HL@box}[2][]{%
  \tikz[#1]{%
    \pgfpathrectangle{\pgfpoint{0.3pt}{0pt}}{\pgfpoint{\wd #2}{\ht #2}}%
    \pgfusepath{use as bounding box}%
    \node[anchor=base west,fill=lightorange,outer sep=0pt,inner xsep=0.3pt,inner ysep=0pt,minimum height=\ht\strutbox+0.3pt,#1]{\raisebox{0.3pt}{\strut}\strut\usebox{#2}};
  }%
}
\makeatother

\usetikzlibrary{calc,trees,positioning,arrows,chains,shapes.geometric,%
  decorations.pathreplacing,decorations.pathmorphing,decorations.text,shapes,%
  matrix,shapes.symbols,patterns,shadows}

\def\addlegendimage{\csname pgfplots@addlegendimage\endcsname}

\pgfplotsset{compat=newest}

\newcommand*\Let[2]{\State #1 $\gets$ #2}
\newcommand*\LetHL[2]{\State {\btHL[fill=light-gray] #1 $\gets$ #2}}
\newcommand*\Fcall[1]{\textsc{#1}}
\newcommand*\IfHL[1]{\State {\btHL[fill=light-gray] \textbf{if} #1 \textbf{then}}}
\newcommand*\ElseHL{\State {\btHL[fill=light-gray] \textbf{else}}}
\algrenewcommand\alglinenumber[1]{\tiny\color{Black!70}{#1}}
\algrenewcommand\algorithmicforall[2]{\textbf{for} $i=$ #1 \textbf{to} #2}
\algrenewtext{ForAll}{\algorithmicforall}
\algnewcommand\algorithmicswitch{\textbf{switch}}
\algnewcommand\algorithmiccase{\textbf{case}}
\algdef{SE}[SWITCH]{Switch}{EndSwitch}[1]{\algorithmicswitch\ #1\ \algorithmicdo}{\algorithmicend\ \algorithmicswitch}%
\algdef{SE}[CASE]{Case}{EndCase}[1]{\algorithmiccase\ #1}{\algorithmicend\ \algorithmiccase}%
\algtext*{EndSwitch}%
\algtext*{EndCase}%
\algdef{SE}[SUBALG]{Indent}{EndIndent}{}{\algorithmicend\ }%
\algtext*{Indent}
\algtext*{EndIndent}

\definecolor{verylightgray}{rgb}{.97,.97,.97}

\lstdefinelanguage{Solidity}{
	keywords=[1]{anonymous, assembly, assert, balance, break, call, callcode, case, catch, class, constant, constructor, continue, contract, debugger, default, delegatecall, delete, do, else, event, export, external, false, finally, for, function, gas, if, implements, import, in, indexed, instanceof, interface, internal, is, length, library, log0, log1, log2, log3, log4, memory, modifier, new, payable, pragma, private, protected, public, pure, push, require, return, returns, revert, selfdestruct, send, storage, struct, suicide, super, switch, then, this, throw, transfer, true, try, typeof, using, value, view, while, with, addmod, ecrecover, keccak256, mulmod, ripemd160, sha256, sha3}, 
	keywordstyle=[1]\color{DarkBlue}\bfseries,
	keywords=[2]{address, bool, byte, bytes, bytes1, bytes2, bytes3, bytes4, bytes5, bytes6, bytes7, bytes8, bytes9, bytes10, bytes11, bytes12, bytes13, bytes14, bytes15, bytes16, bytes17, bytes18, bytes19, bytes20, bytes21, bytes22, bytes23, bytes24, bytes25, bytes26, bytes27, bytes28, bytes29, bytes30, bytes31, bytes32, enum, int, int8, int16, int24, int32, int40, int48, int56, int64, int72, int80, int88, int96, int104, int112, int120, int128, int136, int144, int152, int160, int168, int176, int184, int192, int200, int208, int216, int224, int232, int240, int248, int256, mapping, string, uint, uint8, uint16, uint24, uint32, uint40, uint48, uint56, uint64, uint72, uint80, uint88, uint96, uint104, uint112, uint120, uint128, uint136, uint144, uint152, uint160, uint168, uint176, uint184, uint192, uint200, uint208, uint216, uint224, uint232, uint240, uint248, uint256, var, void, ether, finney, szabo, wei, days, hours, minutes, seconds, weeks, years},	
	keywordstyle=[2]\color{teal}\bfseries,
	keywords=[3]{block, blockhash, coinbase, difficulty, gaslimit, number, timestamp, msg, data, gas, sender, sig, value, now, tx, gasprice, origin},	
	keywordstyle=[3]\color{violet}\bfseries,
        keywords=[4]{minimize},
        keywordstyle=[4]\color{BrickRed}\bfseries,
	identifierstyle=\color{black},
	sensitive=false,
	comment=[l]{//},
	morecomment=[s]{/*}{*/},
	commentstyle=\color{gray}\ttfamily,
	stringstyle=\color{red}\ttfamily,
	morestring=[b]',
	morestring=[b]"
}

\lstdefinestyle{solidity}{
	language=Solidity,
	extendedchars=true,
        upquote          = true,%
	basicstyle=\footnotesize\ttfamily,
        columns          = [c]fixed,%
        aboveskip        = 0mm,%
        belowskip        = 2mm,%
        keepspaces       = true,%
        mathescape       = true,%
	showstringspaces=false,
	showspaces=false,
	numbers=left,
	numberstyle=\tiny\color{Black!70},
	numbersep=4pt,
	tabsize=2,
	breaklines=true,
	showtabs=false,
	captionpos=b,
        escapechar=¤,
        moredelim=**[is][{\btHL[fill=light-gray]}]{°}{°},
        xleftmargin=1.3em,%
}

\lstdefinestyle{basic}{%
  morekeywords     = [1]{var},%
  morekeywords     = [2]{},%
  keywordstyle     = [2]\color{teal}\bfseries,%
  morekeywords     = [3]{minimize},%
  keywordstyle     = [3]\color{BrickRed}\bfseries,%
  keywordstyle     = \bfseries\color{DarkBlue},%
  commentstyle     = \ttfamily\color{Black!70}\lst@ifdisplaystyle\footnotesize\fi,%
  basicstyle       = \ttfamily\lst@ifdisplaystyle\footnotesize\fi,%
  emph             = {int,char,double,float,unsigned,void,bool},%
  emphstyle        = {\color{teal}\bfseries},%
  columns          = [c]fixed,%
  aboveskip        = 0mm,%
  belowskip        = 2mm,%
  keepspaces       = true,%
  mathescape       = true,%
  escapechar       = ¤,%
  tabsize          = 2,%
  numbers          = left,%
  numberstyle      = \tiny\color{Black!70},%
  numbersep        = 4pt,%
  stepnumber       = 1,%
  firstnumber      = 1,%
  showstringspaces = false,%
  captionpos       = b,%
  extendedchars    = true,%
  upquote          = true,%
  abovecaptionskip = 0mm,%
  belowcaptionskip = 0mm,%
  moredelim        = **[is][{\btHL[fill=light-gray]}]{°}{°},%
}

\lstdefinestyle{clang}{%
  language         = C,%
  style            = basic,%
}

\newcommand\code[1]{\lstinline[style=clang]{#1}}

\newcommand\harvey{\textsc{Harvey}\xspace}

\begin{document}

\title{\harvey: A Greybox Fuzzer for Smart Contracts}

\author{\IEEEauthorblockN{Valentin W{\"{u}}stholz}
\IEEEauthorblockA{\textit{ConsenSys Diligence, Germany} \\
valentin.wustholz@consensys.net}
\and
\IEEEauthorblockN{Maria Christakis}
\IEEEauthorblockA{\textit{MPI-SWS, Germany} \\
maria@mpi-sws.org}
}

\maketitle

\begin{abstract}

We present \harvey, an industrial greybox fuzzer for smart contracts,
which are programs managing accounts on a blockchain.

Greybox fuzzing is a lightweight test-generation approach that
effectively detects bugs and security vulnerabilities. However,
greybox fuzzers randomly mutate program inputs to exercise new paths;
this makes it challenging to cover code that is guarded by narrow
checks, which are satisfied by no more than a few input
values. Moreover, most real-world smart contracts transition through
many different states during their lifetime, e.g., for every bid in an
auction. To explore these states and thereby detect deep
vulnerabilities, a greybox fuzzer would need to generate sequences of
contract transactions, e.g., by creating bids from multiple users,
while at the same time keeping the search space and test suite
tractable.

In this experience paper, we explain how \harvey alleviates both
challenges with two key fuzzing techniques and distill the main
lessons learned. First, \harvey extends standard greybox fuzzing with
a method for predicting new inputs that are more likely to cover new
paths or reveal vulnerabilities in smart contracts. Second, it fuzzes
transaction sequences in a targeted and demand-driven way. We have
evaluated our approach on 27 real-world contracts. Our experiments
show that the underlying techniques significantly increase \harvey's
effectiveness in achieving high coverage and detecting
vulnerabilities, in most cases orders-of-magnitude faster; they also
reveal new insights about contract code.
\end{abstract}

\section{Introduction}

Smart contracts are programs that manage crypto-currency accounts on a
blockchain. Reliability of these programs is of critical importance
since bugs may jeopardize digital assets.
Automatic test generation has shown to be an effective approach to
find such vulnerabilities, thereby improving software quality.
In fact, there exists a wide variety of test-generation tools, ranging
from random
testing~\cite{ClaessenHughes2000,CsallnerSmaragdakis2004,PachecoLahiri2007},
over greybox fuzzing~\cite{AFL,LibFuzzer}, to dynamic symbolic
execution~\cite{GodefroidKlarlund2005,CadarEngler2005}.

Random
testing~\cite{ClaessenHughes2000,CsallnerSmaragdakis2004,PachecoLahiri2007}
and blackbox fuzzing~\cite{PeachFuzzer,ZzufFuzzer} generate random
inputs to a program, run the program with these inputs, and check for
bugs. Despite the practicality of these techniques, their
effectiveness, that is, their ability to explore new paths, is
limited. The search space of valid program inputs is typically huge,
and a random exploration can only exercise a small fraction of (mostly
shallow) paths.

At the other end of the spectrum, dynamic symbolic
execution~\cite{GodefroidKlarlund2005,CadarEngler2005} and whitebox
fuzzing~\cite{GodefroidLevin2008,CadarDunbar2008,GaneshLeek2009}
repeatedly run a program, both concretely and symbolically. At
runtime, they collect symbolic constraints on program inputs from
branch statements along the execution path. These constraints are then
appropriately modified and a constraint solver is used to generate new
inputs, thereby steering execution toward another path. Although these
techniques are very effective in covering new paths, they cannot be as
efficient and scalable as other test-generation techniques that do not
spend any time on program analysis and constraint solving.

Greybox fuzzing~\cite{AFL,LibFuzzer} lies in the middle of the
spectrum between performance and effectiveness in discovering new
paths. It does not require program analysis or constraint solving, but
it relies on a lightweight program instrumentation that allows the
fuzzer to tell when an input exercises a new path. In other words, the
instrumentation is useful in computing a unique identifier for each
explored path in the program under test. American Fuzzy Lop
(AFL)~\cite{AFL} is a prominent example of a state-of-the-art greybox
fuzzer that has detected numerous bugs and security
vulnerabilities~\cite{AFL-Bugs}.


In this paper, we present \harvey, the first greybox fuzzer for smart
contracts. We report on our experience in designing \harvey, and in
particular, focus on how to alleviate two key challenges we
encountered when fuzzing real-world contracts. Although the challenges
are not exclusive to our specific application domain, our techniques
are shown to be quite effective for smart contracts and there are
important lessons to be learned from our experiments.

\textbf{Challenge \#1.} Despite the fact that greybox fuzzing strikes
a good balance between performance and effectiveness, the inputs are
still randomly mutated, for instance, by flipping arbitrary
bits. As a result, many generated inputs exercise the same program
paths. To address this problem, there have emerged techniques that
direct greybox fuzzing toward low-frequency
paths~\cite{BoehmePham2016}, vulnerable paths~\cite{RawatJain2017},
deep paths~\cite{SparksEmbleton2007}, or specific sets of program
locations~\cite{BoehmePham2017}.
Such techniques have mostly focused on which seed inputs to prioritize
and which parts of these inputs to mutate.

\textbf{Challenge \#2.} Smart contracts may transition through many
different states during their lifetime, for instance, for every bet in
a gambling game. The same holds for any stateful system that is
invoked repeatedly, such as a web service.
Therefore, detecting vulnerabilities in such programs often requires
generating and fuzzing sequences of invocations that explore the
possible states. For instance, to test a smart contract that
implements a gambling game, a fuzzer would need to automatically
create sequences of bets from multiple players.
However, since the number of possible sequences grows exponentially
with the sequence length, it is difficult to efficiently detect the
few sequences that reveal a bug.

\textbf{Our approach and lessons learned.} To alleviate the first
challenge, we developed a technique that systematically predicts new
inputs for the program under test with the goal of increasing the
performance and effectiveness of greybox fuzzing. In contrast to
existing work in greybox fuzzing, our approach suggests concrete input
values based on information from previous executions, instead of
performing arbitrary mutations. And in contrast to whitebox fuzzing,
our input-prediction mechanism remains particularly lightweight.

Inputs are predicted in a way that aims to direct greybox fuzzing
toward \emph{optimal} executions, for instance, defined as executions
that flip a branch condition in order to increase coverage. Our
technique is parametric in what constitutes an optimal execution, and
in particular, in what properties such an execution needs to satisfy.

More specifically, each program execution is associated with zero or
more \emph{cost metrics}, which are computed automatically. A cost
metric captures how close the execution is to satisfying a given
property at a given program location.
Executions that minimize a cost metric are considered optimal with
respect to that metric.
For example, a cost metric could be defined at each arithmetic
operation in the program such that it is minimized (i.e., becomes
zero) when an execution triggers an arithmetic overflow.
Our technique uses the costs that are computed with cost metrics along
executions of the program to \emph{iteratively} predict inputs leading
to optimal executions.

Our experiments show that \harvey is extremely successful in
predicting inputs that flip a branch condition even in a single
iteration (success rate of 99\%). This suggests \emph{a low complexity
  of branch conditions in real-world smart contracts}.

Although this input-prediction technique is very effective in
practice, it is not sufficient for thoroughly testing a smart contract
and its state space. As a result, \harvey generates, executes, and
fuzzes sequences of transactions, which invoke the contract's
functions. Each of these transactions can have side effects on the
contract's state, which may affect the execution of subsequent
invocations. To alleviate the second challenge of exploring the search
space of all possible sequences, we devised a technique for
demand-driven sequence fuzzing, which avoids generating transaction
sequences when they cannot further increase coverage.

Our experiments show that 74\% of \emph{bugs in real smart contracts
  require generating more than one transaction to be found}. This
highlights the need for techniques like ours that are able to
effectively prune the space of transaction sequences.

In total, we evaluate \harvey on 27 Ethereum smart contracts. Our
fuzzer's underlying techniques significantly increase its
effectiveness in achieving high coverage (by up to 3x) and detecting
vulnerabilities, in most cases orders-of-magnitude faster.

\textbf{Contributions.} We make the following contributions:
\begin{itemize}
\item We present \harvey, the first greybox fuzzer for smart
  contracts, which is being used industrially by one of the largest
  blockchain-security consulting companies.

\item We describe our architecture and two key techniques for
  alleviating the important challenges outlined above.

\item We evaluate our fuzzer on 27 real-world benchmarks and
  demonstrate that the underlying techniques significantly increase
  its effectiveness.

\item We distill the main lessons learned from fuzzing smart-contract
  code.
\end{itemize}


\section{Background}
\label{sect:background}

In this section, we give background on standard greybox fuzzing and
smart contracts.

\subsection{Greybox Fuzzing}
\label{subsect:fuzzing}



Alg.~\ref{alg:greyboxFuzzingWithPrediction} shows how greybox fuzzing
works. (The grey boxes should be ignored for now.) The fuzzer takes as
input the program under test $\mathit{prog}$ and a set of seeds
$S$. It starts by running the program with the seeds, and during each
program execution, the instrumentation is able to capture the path
that is currently being explored and associate it with a unique
identifier $\mathit{PID}$ (line~1). Note that the $\mathit{PIDs}$ data
structure is a key-value store from a $\mathit{PID}$ to an input that
exercises the path associated with $\mathit{PID}$.
Next, an input is selected for mutation (line~3), and it is assigned
an ``energy'' value that denotes how many times it should be fuzzed
(line~5).

The input is mutated (line~12), and the program is run with the new
input (line~13). If the program follows a path that has not been
previously explored, the new input is added to the test suite
(lines~14--15).
The above process is repeated until an exploration bound is reached
(line~2). The fuzzer returns a test suite containing one test for each
explored path.
%

\subsection{Smart Contracts}
\label{subsect:contracts}

Ethereum~\cite{EthereumWhitePaper,Ethereum} is one of the most
popular
blockchain-based~\cite{BlockchainBlueprint,BlockchainTechnology,BlockchainRevolution},
distributed-computing platforms~\cite{BartolettiPompianu2017}. It
supports two kinds of accounts, user and contract accounts,
both of which store a balance, are owned by a user, and publicly
reside on the blockchain.

In contrast to a user account, a contract account is managed through
code that is associated with it. The contract code captures
agreements between users, for example, to encode the rules of an auction.
A contract account also has persistent state where the code may store
data, such as auction bids.

Contract accounts, their code, and persistent state are called
\emph{smart contracts}. Programmers may write the code in several
languages, like Solidity or Vyper, all of which compile to the
Ethereum Virtual Machine (EVM)~\cite{EthereumYellowPaper} bytecode.

To interact with a contract, users issue \emph{transactions} that call
its functions, for instance, to bid in an auction, and are required to
pay a fee for transactions to be executed. This fee is called
\emph{gas} and is roughly proportional to how much code is run.

\section{Overview}
\label{sect:overview}

We now give an overview of our approach focusing on the challenges we
aim to alleviate.

\subsection{Challenge \#1: Random Input Mutations}
\label{subsect:challenge1}

Fig.~\ref{fig:exampleFuzzing} shows a constructed smart-contract
function \code{baz} (in Solidity) that takes as input three
(256-bit) integers \code{a}, \code{b}, and \code{c} and returns an
integer. There are five paths in this function, all of which are
feasible. Each path is denoted by a unique return value. (The grey
boxes should be ignored for now.)

When running AFL, a state-of-the-art greybox fuzzer, on (a C version
of) function \code{baz}, only four out of five paths are explored
within 12h. During this time, greybox fuzzing constructs a test suite
of four inputs, each of which exploring a different path.  The path
with return value 2 remains unexplored even after the fuzzer generates
about 311M different inputs. All but four of these inputs are
discarded as they exercise a path in \code{baz} that has already been
covered by a previous test.

\begin{figure}[t]
\begin{lstlisting}[style=solidity]
function baz(int256 a, int256 b, int256 c)
                          returns (int256) {
  int256 d = b + c;
  °minimize(d < 1 ? 1 - d : 0);°  ¤\label{line:minimize1T}¤
  °minimize(d < 1 ? 0 : d);° ¤\label{line:minimize1F}¤
  if (d < 1) { ¤\label{line:equalsPre1}¤
    °minimize(b < 3 ? 3 - b : 0);° ¤\label{line:minimize2T}¤
    °minimize(b < 3 ? 0 : b - 2);° ¤\label{line:minimize2F}¤
    if (b < 3) { ¤\label{line:equalsPre2}¤
      return 1;
    }
    °minimize(a == 42 ? 1 : 0);° ¤\label{line:minimize3T}¤
    °minimize(a == 42 ? 0 : |a - 42|);° ¤\label{line:minimize3F}¤
    if (a == 42) { ¤\label{line:equals}¤
      return 2; ¤\label{line:ret2}¤
    }
    return 3;
  } else {
    °minimize(c < 42 ? 42 - c : 0);° ¤\label{line:minimize4T}¤
    °minimize(c < 42 ? 0 : c - 41);° ¤\label{line:minimize4F}¤
    if (c < 42) {
      return 4;
    }
    return 5;
  }
}
\end{lstlisting}
\vspace{-1em}
\caption{Example for fuzzing with input prediction.}
\label{fig:exampleFuzzing}
\vspace{-1em}
\end{figure}

The path with return value 2 is not covered because greybox fuzzers
randomly mutate program inputs (line~12 of
Alg.~\ref{alg:greyboxFuzzingWithPrediction}). It is generally
challenging for fuzzers to generate inputs that satisfy ``narrow
checks'', that is, checks that only become true for very few input
values (e.g., line~\ref{line:equals} of
Fig.~\ref{fig:exampleFuzzing}). In this case, the probability that the
fuzzer will generate value 42 for input \code{a} is 1 out of $2^{256}$
for 256-bit integers. Even worse, to cover the path with return value
2 (line~\ref{line:ret2}), the sum of inputs \code{b} and \code{c} also
needs to be less than 1 (line~\ref{line:equalsPre1}) and \code{b} must
be greater than or equal to 3 (line~\ref{line:equalsPre2}).
As a result, several techniques have been proposed to guide greybox
fuzzing to satisfy such narrow checks, e.g., by selectively applying
whitebox fuzzing~\cite{StephensGrosen2016}.

\textbf{Fuzzing with input prediction.}
In contrast, our technique for input prediction is more lightweight,
without requiring any program analysis or constraint solving.
It does, however, require additional instrumentation of the program to
collect more information about its structure than standard greybox
fuzzing, thus making fuzzing a lighter shade of grey.
This information captures the distance from an optimal execution at
various points in the program and is then used to predict inputs that
guide exploration toward optimal executions.

Our fuzzer takes as input a program $\mathit{prog}$ and seeds $S$. It
also requires a partial function $f_{\mathit{cost}}$ that maps
execution states to cost metrics.
When execution of $\mathit{prog}$ reaches a state $s$, the fuzzer
evaluates the cost metric $f_{\mathit{cost}}(s)$.
For example, the grey boxes in Fig.~\ref{fig:exampleFuzzing} define a
function $f_{\mathit{cost}}$ for \code{baz}. Each \code{minimize}
statement specifies a cost metric at the execution state where it is
evaluated.
Note that $f_{\mathit{cost}}$ constitutes a runtime instrumentation of
$\mathit{prog}$---we use \code{minimize} statements only for
illustration. A compile-time instrumentation would increase gas usage
of the contract and potentially lead to false positives when detecting
out-of-gas errors.

The cost metrics of Fig.~\ref{fig:exampleFuzzing} define optimal
executions as those that flip a branch condition. Specifically,
consider an execution along which variable \code{d} evaluates to
0. This execution takes the then-branch of the first if-statement, and
the cost metric defined by the \code{minimize} statement on
line~\ref{line:minimize1T} evaluates to 1. This means that the
distance of the current execution from an execution that exercises the
(implicit) else-branch of the if-statement is 1. Now, consider a
second execution that also takes this then-branch (\code{d} evaluates
to --1). In this case, the cost metric on line~\ref{line:minimize1T}
evaluates to 2, which indicates a greater distance from an execution
that exercises the else-branch.


Based on this information, our input-prediction technique is able to
suggest new inputs that make the execution of \code{baz} take the
else-branch of the first if-statement and minimize the cost metric on
line~\ref{line:minimize1T} (i.e., the cost becomes zero). For
instance, assume that the predicted inputs cause \code{d} to evaluate
to 7. Although the cost metric on line~\ref{line:minimize1T} is now
minimized, the cost metric on line~\ref{line:minimize1F} evaluates to
7, which is the distance of the current execution from an execution
that takes the then-branch.

Similarly, the \code{minimize} statements on
lines~\ref{line:minimize2T}--\ref{line:minimize2F},
\ref{line:minimize3T}--\ref{line:minimize3F}, and
\ref{line:minimize4T}--\ref{line:minimize4F} of
Fig.~\ref{fig:exampleFuzzing} define cost metrics that are minimized
when an execution flips a branch condition in a subsequent
if-statement. This instrumentation aims to maximize path coverage, and
for this reason, an execution can never minimize all cost metrics.
In fact, the fuzzer has achieved full path coverage when the generated
tests cover all feasible combinations of branches in the program; that
is, when they minimize all possible combinations of cost metrics.

The fuzzer does not exclusively rely on prediction to generate program
inputs, for instance, when there are not enough executions from which
to make a good prediction. In the above example, the inputs for the
first two executions (where \code{d} is 0 and --1) are generated by
the fuzzer without prediction. Prediction can only approximate
correlations between inputs and their corresponding costs; therefore,
it is possible that certain predicted inputs do not lead to optimal
executions. In such cases, it is also up to standard fuzzing to
generate inputs that cover any remaining paths.

For the example of Fig.~\ref{fig:exampleFuzzing}, \harvey explores all
five paths within 0.27s and after generating only 372 different
inputs.

\subsection{Challenge \#2: State Space Exploration}
\label{subsect:challenge2}

Fig.~\ref{fig:exampleSequences} shows a simple contract
\code{Foo}. The constructor on line~\ref{line:constructor-foo} initializes
variables \code{x} and \code{y}, which are stored in the persistent
state of the contract. In function \code{Bar}, the failing assertion
(line~\ref{line:fail}) denotes a bug. An assertion violation causes a
transaction to be aborted, and as a result, users lose their gas.
Triggering the bug requires a sequence of at least three transactions,
invoking functions \code{SetY(42)}, \code{CopyY()}, and \code{Bar()}.
(Note that a transaction may directly invoke up to one contract
function.) The assertion violation may also be triggered by calling
\code{IncX} 42 times before invoking \code{Bar}.

\begin{figure}[t]
\begin{lstlisting}[style=solidity]
contract Foo {
  int256 private x;
  int256 private y;

  constructor () public { ¤\label{line:constructor-foo}¤
    x = 0;
    y = 0;
  }

  function Bar() public returns (int256) {
    if (x == 42) {
      assert(false); ¤\label{line:fail}¤
      return 1;
    }
    return 0;
  }

  function SetY(int256 ny) public { y = ny; }

  function IncX() public { x++; }

  function CopyY() public { x = y; }
}
\end{lstlisting}
\vspace{-1em}
\caption{Example for demand-driven sequence fuzzing.}
\label{fig:exampleSequences}
\vspace{-1em}
\end{figure}

There are three ways to test this contract with a standard greybox
fuzzer. First, each function could be fuzzed separately without
considering the persistent variables of the contract as fuzzable
inputs. For example, \code{Bar} would be executed only once---it has
zero fuzzable inputs. No matter the initial values of \code{x} and
\code{y}, the fuzzer would only explore one path in \code{Bar}.

Second, each function could be fuzzed separately while considering the
persistent variables as fuzzable inputs. The fuzzer would then try to
explore both paths in \code{Bar} by generating values for \code{x} and
\code{y}. A problem with this approach is that the probability of
generating value 42 for \code{x} is tiny, as discussed earlier. More
importantly however, this approach might result in false positives
when the persistent state generated by the fuzzer is not reachable
with any sequence of transactions. For example, the contract would
never fail if \code{SetY} ensured that \code{y} is never set to 42 and
\code{IncX} only incremented \code{x} up to 41.

Third, the fuzzer could try to explore all paths in all possible
sequences of transactions up to a bounded length. This, however, means
that a path would span all transactions (instead of a single
function). For example, a transaction invoking \code{Bar} and a
sequence of two transactions invoking \code{CopyY} and \code{Bar}
would exercise two different paths in the contract, even though from
the perspective of \code{Bar} this is not the case. With this
approach, the number of possible sequences grows exponentially in
their length, and so does the number of tests in the test suite.
The larger the test suite, the more difficult it becomes to find a
test that, when fuzzed, leads to the assertion in \code{Foo},
especially within a certain time limit.

We propose a technique for demand-driven sequence fuzzing that
alleviates these limitations. First, it discovers that the only branch
in \code{Foo} that requires more than a single transaction to be
covered is the one leading to the assertion in
\code{Bar}. Consequently, \harvey only generates transaction sequences
whose last transaction invokes \code{Bar}.
Second, our technique aims to increase path coverage only of the
function that is invoked by this last transaction. In other words, the
goal of any previous transactions is to set up the state, and path
identifiers are computed only for the last transaction. Therefore,
reaching the assertion in \code{Bar} by first calling \code{SetY(42)}
and \code{CopyY()} or by invoking \code{IncX()} 42 times both result
in covering the same path of the contract.

\harvey triggers the above assertion violation in about 18s.

\section{Fuzzing with Input Prediction}
\label{sect:inputs}

In this section, we present the technical details of how we extend
greybox fuzzing with input prediction.

\subsection{Algorithm}
\label{subsect:algorithm}

The grey boxes in Alg.~\ref{alg:greyboxFuzzingWithPrediction} indicate
the key differences.
In addition to the program under test $\mathit{prog}$ and a set of
seeds $S$, Alg.~\ref{alg:greyboxFuzzingWithPrediction} takes as input
a partial function $f_{\mathit{cost}}$ that, as explained earlier,
maps execution states to cost metrics. The fuzzer first runs the
program with the seeds, and during each program execution, it
evaluates the cost metric $f_{\mathit{cost}}(\mathit{s})$ for every
encountered execution state $\mathit{s}$ in the domain of
$f_{\mathit{cost}}$ (line~1). Like in standard greybox fuzzing, each
explored path is associated with a unique identifier
$\mathit{PID}$. Note, however, that the $\mathit{PIDs}$ data structure
now maps a $\mathit{PID}$ both to an input that exercises the
corresponding path as well as to a cost vector, which records all
costs computed during execution of the program with this input.
Next, an input is selected for mutation (line~3) and assigned an
energy value (line~5).

The input is mutated (line~12), and the program is run with the new
input (line~13). We assume that the new input differs from the
original input (which was selected for mutation on line~3) by the
value of a single input parameter---an assumption that typically holds
for mutation-based greybox fuzzers. As usual, if the program follows a
path that has not been explored, the new input is added to the test
suite (lines~14--15).

On line~17, the original and the new input are passed to the
prediction component of the fuzzer along with their cost vectors. This
component inspects $input$ and $input'$ to determine the input
parameter by which they differ. Based on the cost vectors, it then
suggests a new value for this input parameter such that one of the
cost metrics is minimized.
In case a new input is predicted, the program is tested with this
input, otherwise the original input is mutated (lines~8--10). The
former happens even if the energy of the original input has run out
(line~7) to ensure that we do not waste predicted inputs.

The above process is repeated until an exploration bound is reached
(line 2), and the fuzzer returns a test suite containing one test
for each program path that has been explored.

\textbf{Example.} In Tab.~\ref{tab:runningExample}, we run our
algorithm on the example of Fig.~\ref{fig:exampleFuzzing} step by
step. The first column of the table shows an identifier for every
generated test, and the second column shows the path that each test
exercises identified by the return value of the program. The
highlighted boxes in this column denote paths that are covered for the
first time, which means that the corresponding tests are added to the
test suite (lines~14--15 of
Alg.~\ref{alg:greyboxFuzzingWithPrediction}). The third column shows
the test identifier from which the value of variable $\mathit{input}$
is selected (line~3 of
Alg.~\ref{alg:greyboxFuzzingWithPrediction}). Note that, according to
the algorithm, $\mathit{input}$ is selected from tests in the test
suite.

\setlength{\textfloatsep}{0pt}
\begin{algorithm}[t]
  \caption{\textbf{Greybox fuzzing with input prediction.}}
  \label{alg:greyboxFuzzingWithPrediction}
  \footnotesize\textbf{Input:} Program $\mathit{prog}$, Seeds $S$, {\btHL[fill=light-gray] Cost function $f_{\mathit{cost}}$}
  \begin{algorithmic}[1]
    \small
    \fontsize{8}{10}
      \Let{$\mathit{PIDs}$}{\Fcall{RunSeeds}$(S, \mathit{prog},$ {\btHL[fill=light-gray] $f_{\mathit{cost}}$}$)$}
      \While{$\neg$\Fcall{Interrupted}()}
        \Let{$\mathit{input},$ {\btHL[fill=light-gray] $\mathit{cost}$}}{\Fcall{PickInput}$(\mathit{PIDs})$}
        \Let{$\mathit{energy}$}{0}
        \Let{$\mathit{maxEnergy}$}{\Fcall{AssignEnergy}$(\mathit{input})$}
        \LetHL{$\mathit{predictedInput}$}{\textbf{nil}}
        \While{$\mathit{energy} < \mathit{maxEnergy}$ {\btHL[fill=light-gray] $\vee \; \mathit{predictedInput} \neq$ \textbf{nil}}}
          \IfHL{$\mathit{predictedInput} \neq$ \textbf{nil}}
            \Indent
            \LetHL{$\mathit{input'}$}{$\mathit{predictedInput}$}
            \LetHL{$\mathit{predictedInput}$}{\textbf{nil}}
            \EndIndent
          \ElseHL
            \Indent
            \Let{$\mathit{input'}$}{\Fcall{FuzzInput}$(\mathit{input})$}
            \EndIndent
          \Let{$\mathit{PID'},$ {\btHL[fill=light-gray] $\mathit{cost'}$}}{\Fcall{Run}$(\mathit{input'}, \mathit{prog},$ {\btHL[fill=light-gray] $f_{\mathit{cost}}$} $)$}
          \If{\Fcall{IsNew}($\mathit{PID'}, \mathit{PIDs}$)}
            \Let{$\mathit{PIDs}$}{\Fcall{Add}$(\mathit{PID'}, \mathit{input'},$ {\btHL[fill=light-gray] $\mathit{cost'},$} $\mathit{PIDs})$}
          \EndIf
          \IfHL{$\mathit{energy} < \mathit{maxEnergy}$ }
            \Indent
            \LetHL{$\mathit{predictedInput}$}{\Fcall{Predict}$(\mathit{input},\mathit{cost},\mathit{input'},\mathit{cost'})$}
            \EndIndent
          \Let{$\mathit{energy}$}{$\mathit{energy} + 1$}
        \EndWhile
      \EndWhile
  \end{algorithmic}
  \footnotesize\textbf{Output:} Test suite \textsc{Inputs}$(\mathit{PIDs})$
\end{algorithm}
\setlength{\textfloatsep}{10pt}

The fourth column shows a new input for the program under test; this
input is either a seed or the value of variable $\mathit{input'}$ in
the algorithm, which is obtained with input prediction (line~9) or
fuzzing (line~12). Each highlighted box in this column denotes a
predicted value.  The fifth column shows the cost vector that is
computed when running the program with the new input of the fourth
column. Note that we show only non-zero costs and that the subscript
of each cost denotes the line number of the corresponding
\code{minimize} statement in Fig.~\ref{fig:exampleFuzzing}. The sixth
column shows which costs (if any) are used to predict a new input, and
the last column shows the current energy value of the algorithm's
$\mathit{input}$ (lines~4 and~18).
For simplicity, we consider $\mathit{maxEnergy}$ of
Alg.~\ref{alg:greyboxFuzzingWithPrediction} (line~5) to always have
value 2 in this example. Our implementation, however, incorporates
an existing energy schedule~\cite{BoehmePham2016}.

We assume that the set of seeds $S$ contains only the random input
$(\text{\code{a}} = -1, \text{\code{b}} = 0, \text{\code{c}} = -5)$
(test \#1 in Tab.~\ref{tab:runningExample}). This input is then fuzzed
to produce $(\text{\code{a}} = -1, \text{\code{b}} = -3,
\text{\code{c}} = -5)$ (test \#2), that is, to produce a new value for
input parameter \code{b}. Our algorithm uses the costs computed with
metric $C_7$ to predict a new value for \code{b}. (We explain how new
values are predicted in the next subsection.) As a result, test \#3
exercises a new path of the program (the one with return value
3). From the cost vectors of tests \#1 and \#3, only the costs
computed with metric $C_4$ may be used to predict another value for
\code{b}; costs $C_7$ and $C_8$ are already zero in one of the two
tests, while metric $C_{13}$ is not reached in test \#1. Even though
the energy of the original input (from test \#1) has run out, the
algorithm still runs the program with the input predicted from the
$C_4$ costs (line 7). This results in covering the path with return
value 4.

Next, we select an input from tests \#1, \#3, or \#4 of the test
suite. Let's assume that the fuzzer picks the input from test \#3 and
mutates the value of input parameter \code{a}. Note that the cost
vectors of tests \#3 and \#5 differ only with respect to the $C_{13}$
costs, which are therefore used to predict a new input for \code{a}. The
new input exercises a new path of the program (the one with return
value 2). At this point, the cost vectors of tests \#3 and \#6 cannot
be used for prediction because the costs are either the same ($C_4$ and
$C_8$) or they are already zero in one of the two tests ($C_{12}$ and
$C_{13}$). Since no input is predicted and the energy of the original
input (from test \#3) has run out, our algorithm selects another input
from the test suite.

\begin{table}[t]
\centering
\scalebox{0.74}{
\begin{tabular}{c|c|c|ccc|c|c|c}
\multirow{2}{*}{\textsc{\textbf{Test}}} & \multirow{2}{*}{\textsc{\textbf{Path}}} & \textsc{\textbf{Input}} & \multicolumn{3}{c|}{\textsc{\textbf{New Input}}} & \multirow{2}{*}{\textsc{\textbf{Costs}}} & \textsc{\textbf{Prediction}} & \multirow{2}{*}{\textsc{\textbf{Energy}}}\\
& & \textsc{\textbf{from Test}} & \texttt{a} & \texttt{b} & \texttt{c} & & \textsc{\textbf{Cost}} &\\
\hline
\multirow{2}{*}{1} & \cellcolor{light-gray} & \multirow{2}{*}{--} & \multirow{2}{*}{$-1$} & \multirow{2}{*}{0} & \multirow{2}{*}{$-5$} & $C_4 = 6$ & \multirow{2}{*}{--} & \multirow{2}{*}{--}\\
& \multirow{-2}{*}{\cellcolor{light-gray} 1} & & & & & $C_7 = 3$ & &\\
\hline
\multirow{2}{*}{2} & \multirow{2}{*}{1} & \multirow{2}{*}{1} & \multirow{2}{*}{$-1$} & \multirow{2}{*}{$-3$} & \multirow{2}{*}{$-5$} & $C_4 = 9$ & \multirow{2}{*}{$C_7$} & \multirow{2}{*}{0}\\
& & & & & & $C_7 = 6$ & &\\
\hline
\multirow{3}{*}{3} & \cellcolor{light-gray} & \multirow{3}{*}{1} & \multirow{3}{*}{$-1$} & \cellcolor{light-gray} & \multirow{3}{*}{$-5$} & $C_4 = 3$ & \multirow{3}{*}{$C_4$} & \multirow{3}{*}{1}\\
& \cellcolor{light-gray} & & & \cellcolor{light-gray} & & $C_8 = 1$ & &\\
& \multirow{-3}{*}{\cellcolor{light-gray} 3} & & & \multirow{-3}{*}{\cellcolor{light-gray} 3} & & $C_{13} = 43$ & &\\
\hline
\multirow{2}{*}{4} & \cellcolor{light-gray} & \multirow{2}{*}{1} & \multirow{2}{*}{$-1$} & \cellcolor{light-gray} & \multirow{2}{*}{$-5$} & $C_5 = 1$ & \multirow{2}{*}{--} & \multirow{2}{*}{2}\\
& \multirow{-2}{*}{\cellcolor{light-gray} 4} & & & \multirow{-2}{*}{\cellcolor{light-gray} 6} & & $C_{19} = 47$ & &\\
\hline
\multirow{3}{*}{5} & \multirow{3}{*}{3} & \multirow{3}{*}{3} & \multirow{3}{*}{7} & \multirow{3}{*}{3} & \multirow{3}{*}{$-5$} & $C_4 = 3$ & \multirow{3}{*}{$C_{13}$} & \multirow{3}{*}{0}\\
& & & & & & $C_8 = 1$ & &\\
& & & & & & $C_{13} = 35$ & &\\
\hline
\multirow{3}{*}{6} & \cellcolor{light-gray} & \multirow{3}{*}{3} & \cellcolor{light-gray} & \multirow{3}{*}{3} & \multirow{3}{*}{$-5$} & $C_4 = 3$ & \multirow{3}{*}{--} & \multirow{3}{*}{1}\\
& \cellcolor{light-gray} & & \cellcolor{light-gray} & & & $C_8 = 1$ & &\\
& \multirow{-3}{*}{\cellcolor{light-gray} 2} & & \multirow{-3}{*}{\cellcolor{light-gray} 42} & & & $C_{12} = 1$ & &\\
\hline
\multirow{2}{*}{7} & \multirow{2}{*}{4} & \multirow{2}{*}{4} & \multirow{2}{*}{$-1$} & \multirow{2}{*}{6} & \multirow{2}{*}{0} & $C_5 = 6$ & \multirow{2}{*}{$C_{19}$} & \multirow{2}{*}{0}\\
& & & & & & $C_{19} = 42$ & &\\
\hline
\multirow{2}{*}{8} & \cellcolor{light-gray} & \multirow{2}{*}{4} & \multirow{2}{*}{$-1$} & \multirow{2}{*}{6} & \cellcolor{light-gray} & $C_5 = 48$ & \multirow{2}{*}{--} & \multirow{2}{*}{1}\\
& \multirow{-2}{*}{\cellcolor{light-gray} 5} & & & & \multirow{-2}{*}{\cellcolor{light-gray} 42} & $C_{20} = 1$ & &\\
\end{tabular}}
\vspace{0em}
\caption{Running Alg.~\ref{alg:greyboxFuzzingWithPrediction} on the
  example of Fig.~\ref{fig:exampleFuzzing}.}
\label{tab:runningExample}
\vspace{-0em}
\end{table}

This time, let's assume that the fuzzer picks the input from test \#4
and mutates the value of input parameter \code{c}. From the cost
vectors of tests \#4 and \#7, it randomly selects the $C_{19}$ costs
for predicting a new value for \code{c}. The predicted input exercises the
fifth path of the program, thus achieving full path coverage of
function \code{baz} by generating only 8 tests.

Note that our algorithm makes several non-systematic choices, which
may be random or based on heuristics, such as when function
\textsc{PickInput} picks an input from the test suite, when
\textsc{FuzzInput} selects which input parameter to fuzz, or when
\textsc{Predict} decides which costs to use for prediction. For
illustrating how the algorithm works, we made ``good'' choices such
that all paths are exercised with a small number of tests. In
practice, the fuzzer achieved full path coverage of function
\code{baz} with 372 tests, instead of 8, as we discussed in
Sect.~\ref{subsect:challenge1}.

\subsection{Input Prediction}
\label{subsect:prediction}


Our algorithm passes to the prediction component the input vectors
$\mathit{input}$ and $\mathit{input'}$ and the corresponding cost
vectors $\mathit{cost}$ and $\mathit{cost'}$ (line~17 of
Alg.~\ref{alg:greyboxFuzzingWithPrediction}). The input vectors differ
by the value of a single input parameter, say $i_0$ and $i_1$. Now,
let us assume that the prediction component selects a cost metric to
minimize and that the costs that have been evaluated using this metric
appear as $c_0$ and $c_1$ in the cost vectors. This means that cost
$c_0$ is associated with input value $i_0$, and $c_1$ with $i_1$.

As an example, let us consider tests \#3 and \#5 from
Tab.~\ref{tab:runningExample}. The input vectors differ by the value
of input parameter \code{a}, so $i_0 = -1$ (value of \code{a} in test
\#3) and $i_1 = 7$ (value of \code{a} in test \#5). The prediction
component chooses to make a prediction based on cost metric $C_{13}$
since the cost vectors of tests \#3 and \#5 differ only with respect
to this metric, so $c_0 = 43$ (value of $C_{13}$ in test \#3) and $c_1
= 35$ (value of $C_{13}$ in test \#5).

Using the two data points $(i_0, c_0)$ and $(i_1, c_1)$, the goal is
to find a value $i$ such that the corresponding cost is zero. In other
words, our technique aims to find a root of the unknown, but
computable, function that relates input parameter \code{a} to cost
metric $C_{13}$. While there is a wide range of root-finding
algorithms, \harvey uses the \emph{Secant method}. Like other methods,
such as Newton's, the Secant method tries to find a root by performing
successive approximations.

Its basic approximation step considers the two data points as
x-y-coordinates on a plane. Our technique then fits a straight line
$c(i) = m * i + k$ through the points, where $m$ is the slope of the
line and $k$ is a constant. To predict the new input value, it
determines the x-coordinate $i$ where the line intersects with the
x-axis (i.e., where the cost is zero).





From the points $(-1, 43)$ and $(7, 35)$ defined by tests \#3 and
\#5, we compute the line to be $c(i) = -i + 42$. Now, for the cost to
be zero, the value of parameter \code{a} must be 42.
Indeed, when \code{a} becomes 42 in test \#6, cost metric $C_{13}$ is
minimized.

This basic approximation step is precise if the target cost metric is
indeed linear (or piece-wise linear) with the input parameter for
which we are making a prediction. If not, the approximation may fail
to minimize the cost metric. In such cases, \harvey applies the basic
step iteratively (as the Secant method). Our experiments show that one
iteration is typically sufficient for real contracts.

\subsection{Cost Metrics}
\label{subsect:costMetrics}

We now describe the different cost metrics that our fuzzer aims to
minimize: (1)~ones that are minimized when execution flips a branch
condition, and (2)~ones that are minimized when execution is able to
modify arbitrary memory locations.

\textbf{Branch conditions.} We have already discussed cost metrics
that are minimized when execution flips a branch condition in the
example of Fig.~\ref{fig:exampleFuzzing}. Here, we show how the cost
metrics are automatically derived from the program under test.

For the comparison operators \code{==} ($\mathit{eq}$), \code{<}
($\mathit{lt}$), and \code{<=} ($\mathit{le}$), we define the
following cost functions:

\scriptsize
\[
\begin{array}{llllll}
  C_{\mathit{eq}}(l, r)  & = & \begin{cases} 1, & l = r\\ 0, & l \neq r \end{cases} &
  C_{\mathit{\overline{eq}}}(l, r)  & = & \begin{cases} 0, & l = r\\ |l - r|, & l \neq r \end{cases}\\
  C_{\mathit{lt}}(l, r)  & = & \begin{cases} r - l, & l < r\\ 0, & l \geq r \end{cases} &
  C_{\mathit{\overline{lt}}}(l, r)  & = & \begin{cases} 0, & l < r\\ l - r + 1, & l \geq r \end{cases}\\
  C_{\mathit{le}}(l, r) & = & \begin{cases} r - l + 1, & l \leq r\\ 0, & l > r \end{cases} &
  C_{\mathit{\overline{le}}}(l, r) & = & \begin{cases} 0, & l \leq r\\ l - r, & l > r \end{cases}
\end{array}
\]
\normalsize

\noindent
Function $C_{\mathit{eq}}$ from above is non-zero when a branch
condition $l$ \code{==} $r$ holds; it defines the cost metric for
making this condition false. On the other hand, function
$C_{\mathit{\overline{eq}}}$ defines the cost metric for making the
same branch condition true. The arguments $l$ and $r$ denote the left
and right operands of the operator. The notation is similar for all
other functions.


Based on these cost functions, our instrumentation evaluates two cost
metrics before every branch condition in the program under test. The
metrics that are evaluated depend on the comparison operator used in
the branch condition. The cost functions for other comparison
operators, i.e., \code{!=} ($\mathit{ne}$), \code{>} ($\mathit{gt}$),
and \code{>=} ($\mathit{ge}$), are easily derived from the functions
above, and our tool supports them.
Note that our implementation works on the bytecode, where
logical operators, like \code{&&} or \code{||}, are expressed as
branch conditions. We, thus, do not define cost functions for
such operators, but they are also straightforward.







Observe that the inputs of the above cost functions are the operands
of comparison operators, and not program inputs. This makes the cost
functions \emph{precise}, that is, when a cost is minimized, the
corresponding branch is definitely flipped. Approximation can only be
introduced when computing the correlation between a program input and
a cost (Sect.~\ref{subsect:prediction}).

\textbf{Memory accesses.} 
%
To illustrate the flexibility of our cost metrics, we now show another
instantiation that targets a vulnerability specific to smart
contracts.
Consider the example in Fig.~\ref{fig:memoryExample}. (The grey box
should be ignored for now.) It is a simplified version of code
submitted to the Underhanded Solidity Coding Contest (USCC) in
2017~\cite{USCC}.
The USCC is a contest to write seemingly harmless
Solidity code that, however, disguises unexpected vulnerabilities.

\begin{figure}[t]
\begin{lstlisting}[style=solidity]
contract Wallet {
  address private owner; ¤\label{line:field1}¤
  uint[] private bonusCodes; ¤\label{line:field2}¤

  constructor() public { ¤\label{line:constructor-wallet}¤
    owner = msg.sender;
    bonusCodes = new uint[](0);
  }

  function() public payable { } ¤\label{line:payable}¤

  function PushCode(uint c) public {
    bonusCodes.push(c);
  }

  function PopCode() public {
    require(0 <= bonusCodes.length); ¤\label{line:precondition1}¤
    bonusCodes.length--; ¤\label{line:overflow}¤
  }

  function SetCodeAt(uint idx, uint c) public {
    require(idx < bonusCodes.length);
    °minimize(|&(bonusCodes[idx]) - 0xffcaffee|);° ¤\label{line:minimize}¤
    bonusCodes[idx] = c; ¤\label{line:store}¤
  }

  function Destroy() public { ¤\label{line:destroy}¤
    require(msg.sender == owner); ¤\label{line:precondition2}¤
    selfdestruct(msg.sender); ¤\label{line:selfdestruct}¤
  }
}
\end{lstlisting}
\vspace{-1em}
\caption{Example of a memory-access vulnerability.}
\label{fig:memoryExample}
\vspace{-0em}
\end{figure}

The contract of Fig.~\ref{fig:memoryExample} implements a wallet that
has an owner and stores an array (with variable length) of bonus codes
(lines~\ref{line:field1}--\ref{line:field2}). The constructor
(line~\ref{line:constructor-wallet}) initializes the owner to the caller's
address and the bonus codes to an empty array. The empty function
(line~\ref{line:payable}) ensures that assets can be payed to the
wallet. The other functions allow bonus codes to be pushed, popped, or
updated. The last function (line~\ref{line:destroy}) must be called
only by the owner and causes the wallet to self-destruct---to
transfer all assets to the owner and destroy itself.

The vulnerability in this code is caused by the precondition on
line~\ref{line:precondition1}, which should require the array length
to be greater than zero (not equal) before popping an element. When
the array is empty, the statement on line~\ref{line:overflow} causes
the (unsigned) array length to underflow; this effectively disables
the bound-checks of the array, allowing elements to be stored anywhere
in the persistent state of the contract. Therefore, by setting a
bonus code at a specific index in the array, an attacker could
overwrite the address of the owner to their own address. Then, by
destroying the wallet, the attacker would transfer all assets to their
account. In a more optimistic scenario, the owner could be
accidentally set to an invalid address, in which case the assets in
the wallet would become inaccessible.

To detect such vulnerabilities, a greybox fuzzer can, for every
assignment to the persistent state of a contract, pick an arbitrary
address and compare it to the target address of the assignment. When
these two addresses happen to be the same, it is very likely that the
assignment may also target other arbitrary addresses, perhaps as a
result of an exploit. A fuzzer without input prediction, however, is
only able to detect these vulnerabilities by chance, and chances are
extremely low that the target address of an assignment matches an
arbitrarily selected address, especially given that these are 32 bytes
long. In fact, when disabling \harvey's input prediction, the
vulnerability in the code of Fig.~\ref{fig:memoryExample} is not
detected within 12h.

To direct the fuzzer toward executions that could reveal such
vulnerabilities, we define the following cost function:

\vspace{-0em}
\scriptsize
\[
  C_{\mathit{st}}(\mathit{lhsAddr}, \mathit{addr}) = |\mathit{lhsAddr} - \mathit{addr}|\\
\]
\normalsize

\noindent
Here, $\mathit{lhsAddr}$ denotes the address of the
left-hand side of an assignment to persistent state (that is,
excluding assignments to local variables) and $\mathit{addr}$ an
arbitrary address. Function $C_{\mathit{st}}$ is non-zero when
$\mathit{lhsAddr}$ and $\mathit{addr}$ are different, and therefore,
optimal executions are those where the assignment writes to the
arbitrary address, potentially revealing a vulnerability.

Our instrumentation evaluates the corresponding cost metric before
every assignment to persistent state in the program under test. An
example is shown on line~\ref{line:minimize} of
Fig.~\ref{fig:memoryExample}. (We use the \code{&} operator to denote
the address of \code{bonusCodes[idx]}, and we do not show the
instrumentation at every assignment to avoid clutter.) Our fuzzer with
input prediction detects the vulnerability in the contract of
Fig.~\ref{fig:memoryExample} within a few seconds.


Detecting such vulnerabilities based on whether an assignment could
target an arbitrary address might generate false positives when the
address is indeed an intended target of the assignment.  However, the
probability of this occurring in practice is extremely low (again due
to the address length). We did not encounter any false positives
during our experiments.

In general, defining other cost functions is straightforward as long
as there is an expressible measure for the distance between a current
execution and an optimal one.

\section{Demand-Driven Sequence Fuzzing}
\label{sect:sequences}

Recall from Sect.~\ref{subsect:challenge2} that \harvey uses
demand-driven sequence fuzzing to set up the persistent state for
testing the last transaction in the sequence. The goal is to explore
new paths in the function that this transaction invokes, and thus,
detect more bugs.
As explained earlier, directly fuzzing the state, for instance,
variables \code{x} and \code{y} of Fig.~\ref{fig:exampleSequences},
might lead to false positives. Nonetheless, \harvey uses this
aggressive approach when fuzzing transaction sequences to determine
whether a different persistent state can increase path coverage.

The key idea is to generate longer transaction sequences on
demand. This is achieved by fuzzing a transaction sequence in two
modes: \emph{regular}, which does not directly fuzz the persistent
state, and \emph{aggressive}, which is enabled with probability 0.125
and may fuzz the persistent state directly. If \harvey is able to
increase coverage of the last transaction in the sequence using the
aggressive mode, the corresponding input is discarded (because it
might lead to false positives), but longer sequences are generated
when running in regular mode in the future.

For instance, when fuzzing a transaction that invokes \code{Bar} from
Fig.~\ref{fig:exampleSequences}, \harvey temporarily considers
\code{x} and \code{y} as fuzzable inputs of the function.  If this
aggressive fuzzing does not discover any more paths, then \harvey does
not generate additional transactions before the invocation of
\code{Bar}. If, however, the aggressive fuzzing does discover new
paths, our tool generates and fuzzes transaction sequences whose last
transaction calls \code{Bar}. That is, longer transaction sequences
are only generated when they might be able to set up the state before
the last transaction such that its coverage is increased.

For our example, \harvey generates the sequence \code{SetY(42)},
\code{CopyY()}, and \code{Bar()} that reaches the assertion in about
18s. At this point, the fuzzer stops exploring longer sequences for
contract \code{Foo} because aggressively fuzzing the state cannot
further increase the already achieved coverage.

We make two important observations. First, \harvey is so quick in
finding the right argument for \code{SetY} due to input
prediction. Second, demand-driven sequence fuzzing relies on path
identifiers to span no more than a single transaction. Otherwise,
aggressive fuzzing would not be able to determine if longer sequences
may increase coverage of the contract.

\textbf{Mutation operations.}
To generate and fuzz sequences of transactions, \harvey applies three
mutation operations to a given transaction $t$: (1)~fuzz transaction
$t$, which fuzzes the inputs of its invocation, (2)~insert a new
transaction before $t$, and (3)~replace the transactions before $t$
with another sequence.

\harvey uses two pools for efficiently generating new transactions or
sequences, respectively. These pools store transactions or sequences
that are found both to increase coverage of the contract under test
and to modify the persistent state in a way that has not been explored
before. \harvey selects new transactions or sequences from these pools
when applying the second and third mutation operations.

\section{Experimental Evaluation}
\label{sect:evaluation}

In this section, we evaluate \harvey on real-world smart
contracts. First, we explain the benchmark selection and setup. We
then compare different \harvey configurations to assess the
effectiveness of our two fuzzing techniques. At the same time, we
highlight key insights about smart-contract code.

\subsection{Benchmark Selection}
\label{sect:benchmark_selection}

We collected all contracts from 17 GitHub repositories. We selected
the repositories based on two main criteria to obtain a diverse set of
benchmarks. On one hand, we picked popular projects in the Ethereum
community (e.g., the Ethereum Name Service auction, the ConsenSys
wallet, and the MicroRaiden payment service) and with high popularity
on GitHub (4'857 stars in total on 2019-05-07, median 132). Most
contracts in these projects have been reviewed by independent auditors
and are deployed on the Ethereum blockchain, managing significant
amounts of crypto-assets on a daily basis. On the other hand, we also
selected repositories from a wide range of application domains (e.g.,
auctions, token sales, payment networks, and wallets) to cover various
features of the EVM and Solidity. We also included contracts that had
been hacked in the past (The DAO and the Parity wallet) and five
contracts (incl. the four top-ranked entries) from the repository of
the USCC to consider some malicious or buggy contracts.

\begin{table}[t!]
\centering
\scalebox{0.83}{
\begin{tabular}{r|l|r|r|l}
\multicolumn{1}{c|}{\textbf{BIDs}} & \multicolumn{1}{c|}{\textbf{Name}} & \multicolumn{1}{c|}{\textbf{Functions}} & \multicolumn{1}{c|}{\textbf{LoSC}} & \multicolumn{1}{c}{\textbf{Description}}\\
\hline
1      & ENS   & 24  & 1205 & ENS domain name auction\\
2--3   & CMSW  & 49  & 503  & ConsenSys multisig wallet\\
4--5   & GMSW  & 49  & 704  & Gnosis multisig wallet\\
6      & BAT   & 23  & 191  & BAT token (advertising)\\
7      & CT    & 12  & 200  & ConsenSys token library\\
8      & ERCF  & 19  & 747  & ERC Fund (investment fund)\\
9      & FBT   & 34  & 385  & FirstBlood token (e-sports)\\
10--13 & HPN   & 173 & 3065 & Havven payment network\\
14     & MR    & 25  & 1053 & MicroRaiden payment service\\
15     & MT    & 38  & 437  & MOD token (supply-chain)\\
16     & PC    & 7   & 69   & Payment channel\\
17--18 & RNTS  & 49  & 749  & Request Network token sale\\
19     & DAO   & 23  & 783  & The DAO organization\\
20     & VT    & 18  & 242  & Valid token (personal data)\\
21     & USCC1 & 4   & 57   & USCC'17 entry\\
22     & USCC2 & 14  & 89   & USCC'17 (honorable mention)\\
23     & USCC3 & 21  & 535  & USCC'17 (3rd place)\\
24     & USCC4 & 7   & 164  & USCC'17 (1st place)\\
25     & USCC5 & 10  & 188  & USCC'17 (2nd place)\\
26     & PW    & 19  & 549  & Parity multisig wallet\\
27     & BNK    & 44  & 649  & Bankera token\\
\hline
\multicolumn{2}{c|}{\textbf{Total}} & 662 & \multicolumn{1}{r}{12564} &
\end{tabular}
}
\vspace{-0em}
\caption{Overview of benchmarks.}
\label{tab:benchmarks}
\vspace{-0em}
\end{table}

From each of the selected repositories, we identified one or more main
contracts that would serve as contracts under test, resulting in a
total of 27 benchmarks. Note that many repositories contain several
contracts (including libraries) to implement a complex system, such as
an auction. Tab.~\ref{tab:benchmarks} gives an overview of all
benchmarks and the projects from which they originate. The first
column lists the benchmark IDs and the second the project name. The
third and fourth columns show the number of public functions and the
lines of Solidity source code (LoSC) in each benchmark. The
appendix provides details about the tested changesets.

To select our benchmarks, we followed published guidelines on
evaluating fuzzers~\cite{KleesRuef2018}. We do not simply scrape
contracts from the blockchain since most are created with no quality
control and many contain duplicates---contracts without assets or
users are essentially dead code. Moreover, good-quality contracts
typically have dependencies (e.g., on libraries or other contracts)
that would likely not be scraped with them.

In terms of size, note that most contracts are a few hundred lines of
code. Nonetheless, they are complex programs, each occupying at least
a couple of auditors for weeks. More importantly, their size does not
necessarily represent how difficult it is for a fuzzer to test all
paths. For instance, Fig.~\ref{fig:exampleFuzzing} is very small, but
AFL fails to cover all paths within 12h.

\subsection{Experimental Setup}
\label{sect:setup}

We ran different configurations of \harvey and compared the achieved
coverage and required time to detect a bug.

For simplicity, our evaluation focuses on detecting two types of
bugs. First, we detect crashes due to assertion violations (SWC-110
according to the Smart Contract Weakness Classification~\cite{SWC});
in addition to user-provided checks, these include checked errors,
such as division by zero or out-of-bounds array access, inserted by
the compiler. At best, these bugs cause a transaction to be aborted
and waste gas fees. In the worst case, they prevent legitimate
transactions from succeeding, putting assets at risk. For instance, a
user may not be able to claim an auctioned item due to an
out-of-bounds error in the code that iterates over an array of bidders
to determine the winner. Second, we detect memory-access errors
(SWC-124~\cite{SWC}) that may allow an attacker to modify the
persistent state of a contract (Fig.~\ref{fig:memoryExample}).
In practice, \harvey covers a wide range of test
oracles\footnote{SWC-101, 104, 107, 110, 123, 124, 127}, such as
reentrancy and overflows.

For bug de-duplication, it uses a simple approach (much more
conservative than AFL): two bugs of the same type are duplicates if
they occur at the same program location.

For each configuration, we performed 24 runs, each with independent
random seeds, an all-zero seed input, and a time limit of one hour; we
report medians unless stated otherwise. In addition, we performed
Wilcoxon-Mann-Whitney U tests to determine if differences in
medians are statistically significant and report the computed
p-values.

We used an Intel\textregistered~Xeon\textregistered~CPU~@~2.90GHz
36-core machine with 60GB running Ubuntu 18.04.

\subsection{Results}
\label{subsect:results}

We assess \harvey's effectiveness by evaluating three research
questions. The first two focus on our input-prediction technique and
the third on demand-driven sequence fuzzing.

Our baselines implement standard greybox fuzzing within \harvey. Since
there are no other greybox fuzzers for smart contracts, we consider
these suitable. Related work (e.g., \cite{JiangLiu2018,LuuChu2016})
either uses fundamentally different bug-finding techniques (like
blackbox fuzzing or symbolic execution) or focuses on detecting
different types of bugs. Our evaluation aims to demonstrate
improvements over standard greybox fuzzing, the benefits of which have
been shown independently.

\begin{table}[t!]
\centering
\scalebox{0.78}{
\begin{tabular}{r|r|r|r|r|r|r|r|r}
\multicolumn{1}{c|}{\textbf{BID}} & \multicolumn{1}{c|}{\textbf{Bug ID}} & \multicolumn{1}{c|}{\textbf{SWC ID}} & \multicolumn{1}{c|}{$\text{\textbf{T}}_{\text{\textbf{A}}}$} & \multicolumn{1}{c|}{$\text{\textbf{T}}_{\text{\textbf{B}}}$} & \multicolumn{1}{c|}{$\text{\textbf{T}}_{\text{\textbf{A}}}/\text{\textbf{T}}_{\text{\textbf{B}}}$} & \multicolumn{1}{c|}{$\text{\textbf{p}}$} & \multicolumn{1}{c|}{$\text{\textbf{A12}}_{\text{\textbf{A}}}$} & \multicolumn{1}{c}{$\text{\textbf{A12}}_{\text{\textbf{B}}}$}\\
\hline
2  & 990d9524 & SWC-110 & 22.27   & \textbf{0.21   } & 107.35 & 0.000 & 0.00 & 1.00\\
2  & b4f9a3d6 & SWC-110 & 41.59   & \textbf{1.27   } & 32.80 & 0.000 & 0.05 & 0.95\\
3  & c56e90ab & SWC-110 & 8.83    & \textbf{0.21   } & 42.40 & 0.000 & 0.00 & 1.00\\
3  & cb2847d0 & SWC-110 & 13.56   & \textbf{0.85   } & 15.95 & 0.000 & 0.04 & 0.96\\
4  & 306fa4fe & SWC-110 & 34.47   & \textbf{2.62   } & 13.16 & 0.000 & 0.04 & 0.96\\
4  & 57c85623 & SWC-110 & 17.34   & \textbf{0.16   } & 106.62 & 0.000 & 0.00 & 1.00\\
5  & 51444152 & SWC-110 & 11.79   & \textbf{0.14   } & 83.23 & 0.000 & 0.00 & 1.00\\
5  & f6ee56cd & SWC-110 & 14.85   & \textbf{1.14   } & 13.09 & 0.000 & 0.06 & 0.94\\
8  & c9c0b2f4 & SWC-110 & 3600.00 & \textbf{90.66  } & 39.71 & 0.000 & 0.00 & 1.00\\
13 & 341911e4 & SWC-110 & 21.27   & \textbf{4.19   } & 5.07 & 0.000 & 0.05 & 0.95\\
13 & 1cd27b5d & SWC-110 & 30.56   & \textbf{4.60   } & 6.65 & 0.000 & 0.05 & 0.95\\
13 & 26aee7ba & SWC-110 & 20.23   & \textbf{4.11   } & 4.92 & 0.000 & 0.05 & 0.95\\
13 & d7d04622 & SWC-110 & 18.42   & \textbf{4.14   } & 4.45 & 0.000 & 0.06 & 0.94\\
15 & dec48390 & SWC-110 & 2823.62 & 1779.98 & 1.59 & 0.362 & 0.42 & 0.58\\
15 & 193c72a2 & SWC-110 & 3600.00 & \textbf{17.65  } & 204.02 & 0.000 & 0.00 & 1.00\\
15 & 7c3dd9f4 & SWC-110 & 3600.00 & \textbf{221.12 } & 16.28 & 0.000 & 0.04 & 0.96\\
15 & 65aa7261 & SWC-110 & 3600.00 & 3600.00 & 1.00 & 0.338 & 0.48 & 0.52\\
17 & 21646ab7 & SWC-110 & 3600.00 & \textbf{273.17 } & 13.18 & 0.000 & 0.02 & 0.98\\
18 & 3021c487 & SWC-110 & 9.59    & \textbf{0.54   } & 17.68 & 0.000 & 0.03 & 0.97\\
18 & ed97030c & SWC-110 & 3600.00 & 3600.00 & 1.00 & 0.917 & 0.51 & 0.49\\
19 & e3468a11 & SWC-110 & 7.98    & \textbf{0.12   } & 64.62 & 0.000 & 0.00 & 1.00\\
19 & b359efbc & SWC-110 & 8.63    & \textbf{0.09   } & 94.28 & 0.000 & 0.01 & 0.99\\
19 & 9e65397d & SWC-110 & 22.86   & \textbf{0.47   } & 48.96 & 0.000 & 0.01 & 0.99\\
19 & 4063c80f & SWC-110 & 20.45   & \textbf{0.46   } & 44.45 & 0.000 & 0.00 & 1.00\\
19 & 49e4a70e & SWC-110 & 55.38   & \textbf{2.55   } & 21.70 & 0.000 & 0.05 & 0.95\\
19 & ee609ac1 & SWC-110 & 16.13   & \textbf{0.71   } & 22.73 & 0.000 & 0.00 & 1.00\\
19 & 21f5c23f & SWC-110 & 23.72   & \textbf{2.52   } & 9.40 & 0.000 & 0.07 & 0.93\\
22 & f3bf5e12 & SWC-110 & 13.13   & \textbf{0.40   } & 33.05 & 0.000 & 0.07 & 0.93\\
22 & 577a74af & SWC-124 & 3600.00 & \textbf{899.60 } & 4.00 & 0.000 & 0.04 & 0.96\\
23 & 1c8acd5e & SWC-110 & 3600.00 & \textbf{193.66 } & 18.59 & 0.000 & 0.00 & 1.00\\
23 & fda2cafa & SWC-110 & 3600.00 & \textbf{218.57 } & 16.47 & 0.000 & 0.00 & 1.00\\
24 & c837a34b & SWC-110 & 1.85    & \textbf{0.04   } & 42.33 & 0.000 & 0.01 & 0.99\\
24 & d602954b & SWC-110 & 3.97    & \textbf{0.12   } & 34.30 & 0.000 & 0.01 & 0.99\\
24 & 863f9452 & SWC-110 & 4.76    & \textbf{0.18   } & 25.96 & 0.000 & 0.04 & 0.96\\
24 & 9774d846 & SWC-110 & 14.41   & \textbf{0.43   } & 33.43 & 0.000 & 0.05 & 0.95\\
24 & 123bf172 & SWC-110 & 238.63  & \textbf{3.01   } & 79.22 & 0.000 & 0.04 & 0.96\\
24 & a97971ca & SWC-110 & 145.79  & \textbf{4.43   } & 32.90 & 0.000 & 0.05 & 0.95\\
24 & 9a771b96 & SWC-110 & 69.35   & \textbf{3.62   } & 19.14 & 0.000 & 0.03 & 0.97\\
24 & dc7bf682 & SWC-110 & 3600.00 & \textbf{0.69   } & 5246.66 & 0.000 & 0.00 & 1.00\\
26 & ccf7bc67 & SWC-110 & 61.42   & \textbf{1.68   } & 36.59 & 0.000 & 0.01 & 0.99\\
27 & f1c8e169 & SWC-110 & 112.98  & \textbf{16.26  } & 6.95 & 0.000 & 0.16 & 0.84\\
27 & 44312719 & SWC-110 & 77.18   & \textbf{1.50   } & 51.36 & 0.000 & 0.01 & 0.99\\
27 & 33c32ef9 & SWC-110 & 60.52   & \textbf{1.08   } & 56.25 & 0.000 & 0.07 & 0.93\\
27 & d499f535 & SWC-110 & 3600.00 & \textbf{7.36   } & 489.38 & 0.000 & 0.00 & 1.00\\
27 & 4fb4fa53 & SWC-110 & 3600.00 & \textbf{67.29  } & 53.50 & 0.000 & 0.00 & 1.00\\
27 & 47c60a93 & SWC-110 & 3600.00 & \textbf{141.16 } & 25.50 & 0.000 & 0.02 & 0.98\\
27 & 6f92fdea & SWC-110 & 3600.00 & 3600.00 & 1.00 & 0.041 & 0.42 & 0.58\\
\hline
\multicolumn{3}{c|}{\textbf{Median}} & 41.59 & \textbf{2.55} & 25.96 & \multicolumn{3}{c}{}\\



\end{tabular}
}
\vspace{-0em}
\caption{Comparing time-to-bug between configuration A (w/o input
  prediction) and B (w/ input prediction).}
\label{tab:resultsAvsB}
\vspace{-0em}
\end{table}

\textbf{RQ1: Effectiveness of input prediction.} To evaluate input
prediction, we compare with a baseline (configuration A), which only
disables prediction. The first column of Tab.~\ref{tab:resultsAvsB}
identifies the benchmark, the second the bug, and the third the bug
type according to the SWC (110 stands for assertion violations and 124
for memory-access errors).

The fourth and fifth columns show the median time (in secs) after
which each unique bug was found by configurations A and B within the
time limit---B differs from A only by enabling input prediction.
\emph{Configuration B finds 43 out of 47 bugs significantly faster
  than A.}
We report the speed-up factor in the sixth column and the significance
level, i.e., p-value, in the seventh (we use $p < 0.05$). As shown in
the table, \emph{configuration B is faster than A by a factor of up to
  5'247 (median 25.96).}
The last two columns compute the Vargha-Delaney A12 effect
sizes~\cite{VarghaDelaney2000}. Intuitively, these show the
probability of configuration A being faster than B and vice versa.
Note that, to compute the median time, we conservatively counted
3'600s for a given run even if the bug was not found. However,
\emph{on average, B detects 10 more bugs.}


Tab.~\ref{tab:resultsAvsBCov} compares A and B with respect to
instruction coverage.  \emph{For 23 out of 27 benchmarks, B achieves
  significantly higher coverage.} The results for path coverage are
very similar.

\vspace{0.75em}
\fbox{\begin{minipage}{0.42\textwidth}

    Input prediction is very effective in both detecting bugs faster
    and achieving higher coverage.

\end{minipage}}
\vspace{0.75em}

\textbf{RQ2: Effectiveness of iterative input prediction.}
Configuration C differs from B in that it does not iteratively apply
the basic approximation step of the Secant method in case it fails to
minimize a cost metric.
For artificial examples with non-linear branch conditions (e.g.,
\texttt{a\textasciicircum4 + a\textasciicircum2 == 228901770}), we
were able to show that this configuration is less efficient than B in
finding bugs.
However, for our benchmarks, there were no significant time
differences between B and C for detecting 45 of 47 bugs.
Similarly, there were no significant differences in instruction
coverage.

During our experiments with C, \emph{we measured the success rate of
  one-shot cost minimization to range between 97\% and 100\% (median
  99\%).} This suggests that complex branch conditions are not very
common in real-world smart contracts.

\vspace{0.75em}
\fbox{\begin{minipage}{0.42\textwidth}

    Even one iteration of the Secant method is extremely
    successful in predicting inputs. This suggests that the vast
    majority of branch conditions are linear (or piece-wise linear)
    with respect to the program inputs.

\end{minipage}}
\vspace{0.75em}



\begin{table}[t!]
\centering
\scalebox{0.78}{
\begin{tabular}{r|r|r|r|r|r|r}
\multicolumn{1}{c|}{\textbf{BID}} & \multicolumn{1}{c|}{$\text{\textbf{C}}_{\text{\textbf{A}}}$} & \multicolumn{1}{c|}{$\text{\textbf{C}}_{\text{\textbf{B}}}$} & \multicolumn{1}{c|}{$\text{\textbf{C}}_{\text{\textbf{B}}}/\text{\textbf{C}}_{\text{\textbf{A}}}$} & \multicolumn{1}{c|}{$\text{\textbf{p}}$} & \multicolumn{1}{c|}{$\text{\textbf{A12}}_{\text{\textbf{A}}}$} & \multicolumn{1}{c}{$\text{\textbf{A12}}_{\text{\textbf{B}}}$}\\
\hline
1  & 3868.00 & 3868.00 & 1.00 & 0.010 & 0.38 & 0.62\\
2  & 3064.00 & \textbf{4005.50} & 1.31 & 0.000 & 0.00 & 1.00\\
3  & 2575.00 & \textbf{3487.00} & 1.35 & 0.000 & 0.00 & 1.00\\
4  & 2791.00 & \textbf{3773.00} & 1.35 & 0.000 & 0.00 & 1.00\\
5  & 2567.00 & \textbf{3501.00} & 1.36 & 0.000 & 0.00 & 1.00\\
6  & 1832.00 & \textbf{1949.00} & 1.06 & 0.000 & 0.00 & 1.00\\
7  & 1524.00 & 1524.00 & 1.00 & 0.000 & 0.50 & 0.50\\
8  & 1051.00 & \textbf{2205.00} & 2.10 & 0.000 & 0.00 & 1.00\\
9  & 2694.00 & \textbf{3468.00} & 1.29 & 0.000 & 0.00 & 1.00\\
10 & 6833.00 & \textbf{7360.50} & 1.08 & 0.000 & 0.00 & 1.00\\
11 & 7295.00 & \textbf{8716.00} & 1.19 & 0.000 & 0.00 & 1.00\\
12 & 2816.00 & \textbf{5165.00} & 1.83 & 0.000 & 0.00 & 1.00\\
13 & 1585.00 & \textbf{4510.00} & 2.85 & 0.000 & 0.00 & 1.00\\
14 & 3822.00 & \textbf{4655.00} & 1.22 & 0.000 & 0.00 & 1.00\\
15 & 3489.00 & \textbf{5078.50} & 1.46 & 0.000 & 0.00 & 1.00\\
16 & 496.00  & 496.00 & 1.00 & 0.000 & 0.50 & 0.50\\
17 & 1832.00 & \textbf{2754.00} & 1.50 & 0.000 & 0.00 & 1.00\\
18 & 2766.00 & \textbf{2930.00} & 1.06 & 0.000 & 0.00 & 1.00\\
19 & 2411.00 & \textbf{2611.00} & 1.08 & 0.000 & 0.00 & 1.00\\
20 & 1635.00 & \textbf{3018.00} & 1.85 & 0.000 & 0.00 & 1.00\\
21 & 349.00  & \textbf{434.00 } & 1.24 & 0.000 & 0.00 & 1.00\\
22 & 919.00  & \textbf{1274.00} & 1.39 & 0.000 & 0.00 & 1.00\\
23 & 1344.00 & \textbf{2095.00} & 1.56 & 0.000 & 0.00 & 1.00\\
24 & 687.00  & \textbf{754.00 } & 1.10 & 0.001 & 0.22 & 0.78\\
25 & 1082.00 & \textbf{1192.00} & 1.10 & 0.000 & 0.00 & 1.00\\
26 & 1606.00 & 1606.00 & 1.00 & 0.000 & 0.50 & 0.50\\
27 & 4232.00 & \textbf{5499.50} & 1.30 & 0.000 & 0.00 & 1.00\\
\hline
\multicolumn{1}{c|}{\textbf{Median}} & 2411.00 & \textbf{3018.00} & 1.29 & \multicolumn{3}{c}{}\\



\end{tabular}
}
\vspace{-0em}
\caption{Comparing instruction coverage for configurations A (w/o
  input prediction) and B (w/ input prediction).}
\label{tab:resultsAvsBCov}
\vspace{-0em}
\end{table}

\textbf{RQ3: Effectiveness of demand-driven sequence fuzzing.}
To evaluate this research question, we compare configuration A with D,
which differs from A by disabling demand-driven sequence fuzzing. In
particular, D tries to eagerly explore all paths in all possible
transaction sequences, where paths span all transactions.
Tab.~\ref{tab:resultsAvsE} shows a comparison between A and D with
respect to time-to-bug for bugs that were found by at least one
configuration. As shown in the table, \emph{A is significantly faster
  than D in detecting 7 out of 35 bugs, with a speed-up of up to
  20x}. Note that all 7 bugs require more than a single transaction to
be detected.
Instruction coverage is very similar between A and D (slightly higher
for A), but A achieves it within a fraction of the time for 19 out of
27 benchmarks.

In total, 26 out of 35 bugs require more than one transaction to be
found. This suggests that real contracts need to be tested with
sequences of transactions, and consequently, there is much to be
gained from pruning techniques like ours.
Our experiments with D also confirm that, when paths span all
transactions, the test suite becomes orders-of-magnitude larger.

\vspace{0.75em}
\fbox{\begin{minipage}{0.42\textwidth}

    Demand-driven sequence fuzzing is effective in pruning the search
    space of transaction sequences; as a result, it detects bugs and
    achieves coverage faster. Such techniques are useful since most
    bugs require invoking multiple transactions to be revealed.

\end{minipage}}
\vspace{0.75em}

\begin{table}[t!]
\centering
\scalebox{0.78}{
\begin{tabular}{r|r|r|r|r|r|r|r|r}
\multicolumn{1}{c|}{\textbf{BID}} & \multicolumn{1}{c|}{\textbf{Bug ID}} & \multicolumn{1}{c|}{\textbf{SWC ID}} & \multicolumn{1}{c|}{$\text{\textbf{T}}_{\text{\textbf{A}}}$} & \multicolumn{1}{c|}{$\text{\textbf{T}}_{\text{\textbf{D}}}$} & \multicolumn{1}{c|}{$\text{\textbf{T}}_{\text{\textbf{A}}}/\text{\textbf{T}}_{\text{\textbf{D}}}$} & \multicolumn{1}{c|}{$\text{\textbf{p}}$} & \multicolumn{1}{c|}{$\text{\textbf{A12}}_{\text{\textbf{A}}}$} & \multicolumn{1}{c}{$\text{\textbf{A12}}_{\text{\textbf{D}}}$}\\
\hline
2  & 990d9524 & SWC-110 & 22.27   & 15.85   & 1.41 & 0.415 & 0.43 & 0.57\\
2  & b4f9a3d6 & SWC-110 & 41.59   & 66.31   & 0.63 & 0.529 & 0.55 & 0.45\\
3  & c56e90ab & SWC-110 & 8.83    & 8.63    & 1.02 & 0.643 & 0.46 & 0.54\\
3  & cb2847d0 & SWC-110 & 13.56   & 12.72   & 1.07 & 0.749 & 0.53 & 0.47\\
4  & 306fa4fe & SWC-110 & 34.47   & 59.90   & 0.58 & 0.477 & 0.56 & 0.44\\
4  & 57c85623 & SWC-110 & 17.34   & 14.96   & 1.16 & 0.942 & 0.49 & 0.51\\
5  & 51444152 & SWC-110 & 11.79   & 12.01   & 0.98 & 0.585 & 0.55 & 0.45\\
5  & f6ee56cd & SWC-110 & 14.85   & 17.69   & 0.84 & 0.529 & 0.55 & 0.45\\
13 & 341911e4 & SWC-110 & 21.27   & 25.28   & 0.84 & 0.749 & 0.53 & 0.47\\
13 & 1cd27b5d & SWC-110 & 30.56   & 28.54   & 1.07 & 0.359 & 0.58 & 0.42\\
13 & 26aee7ba & SWC-110 & 20.23   & 23.12   & 0.87 & 0.571 & 0.55 & 0.45\\
13 & d7d04622 & SWC-110 & 18.42   & 19.05   & 0.97 & 0.942 & 0.51 & 0.49\\
15 & dec48390 & SWC-110 & \textbf{2823.62} & 3600.00 & 0.78 & 0.000 & 0.84 & 0.16\\
18 & 3021c487 & SWC-110 & 9.59    & 9.84    & 0.97 & 0.338 & 0.58 & 0.42\\
18 & ed97030c & SWC-110 & 3600.00 & 3600.00 & 1.00 & 0.028 & 0.65 & 0.35\\
19 & e3468a11 & SWC-110 & 7.98    & 5.96    & 1.34 & 0.439 & 0.43 & 0.57\\
19 & b359efbc & SWC-110 & 8.63    & 8.25    & 1.05 & 0.288 & 0.41 & 0.59\\
19 & 9e65397d & SWC-110 & 22.86   & 22.06   & 1.04 & 0.781 & 0.52 & 0.48\\
19 & 4063c80f & SWC-110 & 20.45   & 15.82   & 1.29 & 0.177 & 0.39 & 0.61\\
19 & 49e4a70e & SWC-110 & 55.38   & 55.67   & 0.99 & 0.585 & 0.55 & 0.45\\
19 & ee609ac1 & SWC-110 & 16.13   & 19.45   & 0.83 & 0.877 & 0.51 & 0.49\\
19 & 21f5c23f & SWC-110 & 23.72   & 16.43   & 1.44 & 0.718 & 0.47 & 0.53\\
22 & f3bf5e12 & SWC-110 & 13.13   & 9.46    & 1.39 & 0.718 & 0.47 & 0.53\\
24 & c837a34b & SWC-110 & 1.85    & 1.98    & 0.94 & 0.673 & 0.54 & 0.46\\
24 & d602954b & SWC-110 & 3.97    & 4.56    & 0.87 & 0.877 & 0.51 & 0.49\\
24 & 863f9452 & SWC-110 & 4.76    & 4.46    & 1.07 & 0.959 & 0.51 & 0.49\\
24 & 9774d846 & SWC-110 & 14.41   & 12.72   & 1.13 & 0.845 & 0.48 & 0.52\\
24 & 123bf172 & SWC-110 & \textbf{238.63}  & 3600.00 & 0.07 & 0.001 & 0.77 & 0.23\\
24 & a97971ca & SWC-110 & \textbf{145.79}  & 2946.47 & 0.05 & 0.005 & 0.74 & 0.26\\
24 & 9a771b96 & SWC-110 & \textbf{69.35}   & 1087.31 & 0.06 & 0.010 & 0.72 & 0.28\\
26 & ccf7bc67 & SWC-110 & \textbf{61.42}   & 426.17  & 0.14 & 0.000 & 0.80 & 0.20\\
27 & f1c8e169 & SWC-110 & \textbf{112.98}  & 504.67  & 0.22 & 0.002 & 0.76 & 0.24\\
27 & 44312719 & SWC-110 & \textbf{77.18}   & 522.75  & 0.15 & 0.018 & 0.70 & 0.30\\
27 & 33c32ef9 & SWC-110 & 60.52   & 83.35   & 0.73 & 0.464 & 0.56 & 0.44\\
27 & 47c60a93 & SWC-110 & 3600.00 & 3600.00 & 1.00 & 0.026 & 0.65 & 0.35\\
\hline
\multicolumn{3}{c|}{\textbf{Median}} & 21.27 & \textbf{19.45} & 0.97 & \multicolumn{3}{c}{}\\



\end{tabular}
}
\vspace{-0em}
\caption{Comparing time-to-bug between configuration A (w/
  demand-driven sequence fuzzing) and D (w/o demand-driven sequence
  fuzzing).}
\label{tab:resultsAvsE}
\vspace{-0em}
\end{table}

\subsection{Threats to Validity}
\label{subsect:threats}

\textbf{External validity.} Our results may not generalize to all
smart contracts or program types~\cite{SiegmundSiegmund2015}. However,
we evaluated our technique on a diverse set of contracts from a wide
range of domains. We, thus, believe that our selection significantly
helps to ensure generalizability. To further improve external
validity, we also provide the versions of all contracts in the
appendix.
Moreover, our comparisons focus on a single fuzzer---we discuss this
below.

\textbf{Internal validity.} Another potential issue has to do with
whether systematic errors are introduced in the
setup~\cite{SiegmundSiegmund2015}. When comparing configurations, we
always used the same seed inputs in order to avoid bias in the
exploration.

\textbf{Construct validity.} Construct validity ensures that the
evaluation measures what it claims.
We compare several configurations of \harvey, and thus, ensure that
any improvements are exclusively due to techniques enabled in a given
configuration.

\section{Related Work}
\label{sect:relatedWork}

\harvey is the first greybox fuzzer for smart contracts. It
incorporates two key techniques, input prediction and demand-driven
sequence fuzzing, that improve its effectiveness.

\textbf{Greybox fuzzing.} There are several techniques that aim to
direct greybox fuzzing toward certain parts of the search space, such
as low-frequency paths~\cite{BoehmePham2016}, vulnerable
paths~\cite{RawatJain2017}, deep paths~\cite{SparksEmbleton2007}, or
specific sets of program locations~\cite{BoehmePham2017}. There are
also techniques that boost fuzzing by smartly selecting and mutating
inputs~\cite{WooCha2013,RebertCha2014,ChaWoo2015}, or by searching for
new inputs using iterative optimization algorithms, such as gradient
descent~\cite{ChenChen2018}, with the goal of increasing branch
coverage.

In general, input prediction could be used in combination with these
techniques. In comparison, our approach predicts concrete input values
based on two previous executions. To achieve this, we rely on
additional, but still lightweight, instrumentation.

\textbf{Whitebox fuzzing.}
%
Whitebox fuzzing is implemented in many tools, such as
EXE~\cite{CadarGanesh2006}, jCUTE~\cite{SenAgha2006},
Pex~\cite{TillmanndeHalleux2008}, BitBlaze~\cite{SongBrumley2008},
Apollo~\cite{ArtziKiezun2010}, S2E~\cite{ChipounovKuznetsov2011}, and
Mayhem~\cite{ChaAvgerinos2012}, and comes in different flavors, such
as probabilistic symbolic execution~\cite{GeldenhuysDwyer2012} or
model-based whitebox fuzzing~\cite{PhamBoehme2016}.

As discussed earlier, our input-prediction technique does not rely on
any program analysis or constraint solving, and our instrumentation is
more lightweight, for instance, we do not keep track of a symbolic
store and path constraints.

\textbf{Hybrid fuzzing.} Hybrid fuzzers combine fuzzing with
other techniques to join their benefits and achieve better
results. For example, Dowser~\cite{HallerSlowinska2013} uses static
analysis to identify code regions with potential buffer
overflows. Similarly, BuzzFuzz~\cite{GaneshLeek2009} uses taint
tracking to discover which input bytes flow to ``attack
points''. Hybrid Fuzz Testing~\cite{Pak2012} first runs symbolic
execution to find inputs that lead to ``frontier nodes'' and then
applies fuzzing on these inputs. On the other hand,
Driller~\cite{StephensGrosen2016} starts with fuzzing and uses
symbolic execution when it needs help in generating inputs that
satisfy complex checks.

In contrast, input prediction extends greybox fuzzing without relying
on static analysis or whitebox fuzzing. \harvey could, however,
benefit from hybrid-fuzzing approaches.

\textbf{Optimization in testing.} Miller and
Spooner~\cite{MillerSpooner1976} were the first to use optimization
methods in generating test data, and in particular, floating-point
inputs. It was not until 1990 that these ideas were extended by Korel
for Pascal programs~\cite{Korel1990}. Such optimization methods have
recently been picked up again~\cite{McMinn2004}, enhanced, and
implemented in various testing tools, such as
FloPSy~\cite{LakhotiaTillmann2010}, CORAL~\cite{SouzaBorges2011},
EvoSuite~\cite{FraserArcuri2011}, AUSTIN~\cite{LakhotiaHarman2013},
CoverMe~\cite{FuSu2017}, and Angora~\cite{ChenChen2018}.
Most of these tools use fitness functions to determine the distance
from a target and attempt to minimize them. For instance, Korel uses
fitness functions that are similar to our cost metrics for flipping
branch conditions. The search is typically iterative, e.g., by using
hill climbing, simulated annealing, or genetic
algorithms~\cite{MetropolisRosenbluth1953,KirkpatrickGelatt1983,PargasHarrold1999}.

Our prediction technique is inspired by these approaches but is
applied in the context of greybox fuzzing. When failing to minimize a
cost metric, \harvey falls back on standard greybox fuzzing, which is
known for its effectiveness. We show that input prediction works
particularly well in the domain of smart contracts, where even one
iteration is generally enough for minimizing a cost metric.

\textbf{Method-call sequence generation.} For testing object-oriented
programs, it is often necessary to generate complex input objects
using sequences of method calls. There are many
approaches~\cite{Tonella2004,XieMarinov2005,InkumsahXie2007,PachecoLahiri2007,ThummalapentaXie2009,ThummalapentaXie2011,ZhangSaff2011,GargIvancic2013}
that automatically generate such sequences using techniques such as
dynamic inference, static analysis, or evolutionary testing.

In contrast, demand-driven sequence fuzzing only relies on greybox
fuzzing and targets smart contracts.

\textbf{Program analysis for smart contracts.} There exist various
applications of program analysis to smart contracts, such as symbolic
execution, static analysis, and
verification~\cite{LuuChu2016,BhargavanDelignat-Lavaud2016,AtzeiBartoletti2017,ChenLi2017,SergeyHobor2017,ChatterjeeGoharshady2018,AmaniBegel2018,BrentJurisevic2018,GrechKong2018,GrossmanAbraham2018,KalraGoel2018,NikolicKolluri2018,TsankovDan2018,Manticore,Mythril}.
The work most closely related to ours is the blackbox fuzzer
ContractFuzzer~\cite{JiangLiu2018} and the property-based testing tool
Echidna~\cite{Echidna}.
In contrast, our technique is the first to apply greybox fuzzing to
smart contracts.

\section{Conclusion}
\label{sect:conclusion}

We presented \harvey, an industrial greybox fuzzer for smart
contracts. During its development, we encountered two key challenges
that we alleviate with input prediction and demand-driven sequence
fuzzing. Our experiments show that both techniques significantly
improve \harvey's effectiveness and highlight certain insights about
contract code.

In future work, we plan to further enhance \harvey by leveraging
complementary techniques, such as static analysis and lightweight
dynamic symbolic execution.


\newpage

\bibliographystyle{IEEEtran}
\bibliography{tandem}

\newpage

\onecolumn

\appendix

\section{Smart Contract Repositories}
\label{sect:repos}

All tested smart contracts are open source. Tab.~\ref{tab:repos}
provides the changeset IDs and links to their repositories.

\begin{table*}[b!]
\centering
\scalebox{0.9}{
\begin{tabular}{r|l|l|l}
\multicolumn{1}{c|}{\textbf{BIDs}} & \multicolumn{1}{c|}{\textbf{Name}} & \multicolumn{1}{c|}{\textbf{Changeset ID}} & \multicolumn{1}{c}{\textbf{Repository}}\\
\hline
1      & ENS   & 5108f51d656f201dc0054e55f5fd000d00ef9ef3 & \url{https://github.com/ethereum/ens}\\
2--3   & CMSW  & 2582787a14dd861b51df6f815fab122ff51fb574 & \url{https://github.com/ConsenSys/MultiSigWallet}\\
4--5   & GMSW  & 8ac8ba7effe6c3845719e480defb5f2ecafd2fd4 & \url{https://github.com/gnosis/MultiSigWallet}\\
6      & BAT   & 15bebdc0642dac614d56709477c7c31d5c993ae1 & \url{https://github.com/brave-intl/basic-attention-token-crowdsale}\\
7      & CT    & 1f62e1ba3bf32dc22fe2de94a9ee486d667edef2 & \url{https://github.com/ConsenSys/Tokens}\\
8      & ERCF  & c7d025220a1388326b926d8983e47184e249d979 & \url{https://github.com/ScJa/ercfund}\\
9      & FBT   & ae71053e0656b0ceba7e229e1d67c09f271191dc & \url{https://github.com/Firstbloodio/token}\\
10--13 & HPN   & 540006e0e2e5ef729482ad8bebcf7eafcd5198c2 & \url{https://github.com/Havven/havven}\\
14     & MR    & 527eb90c614ff4178b269d48ea063eb49ee0f254 & \url{https://github.com/raiden-network/microraiden}\\
15     & MT    & 7009cc95affa5a2a41a013b85903b14602c25b4f & \url{https://github.com/modum-io/tokenapp-smartcontract}\\
16     & PC    & 515c1b935ac43afc6bf54fcaff68cf8521595b0b & \url{https://github.com/mattdf/payment-channel}\\
17--18 & RNTS  & 6c39082eff65b2d3035a89a3f3dd94bde6cca60f & \url{https://github.com/RequestNetwork/RequestTokenSale}\\
19     & DAO   & f347c0e177edcfd99d64fe589d236754fa375658 & \url{https://github.com/slockit/DAO}\\
20     & VT    & 30ede971bb682f245e5be11f544e305ef033a765 & \url{https://github.com/valid-global/token}\\
21     & USCC1 & 3b26643a85d182a9b8f0b6fe8c1153f3bd510a96 & \url{https://github.com/Arachnid/uscc}\\
22     & USCC2 & 3b26643a85d182a9b8f0b6fe8c1153f3bd510a96 & \url{https://github.com/Arachnid/uscc}\\
23     & USCC3 & 3b26643a85d182a9b8f0b6fe8c1153f3bd510a96 & \url{https://github.com/Arachnid/uscc}\\
24     & USCC4 & 3b26643a85d182a9b8f0b6fe8c1153f3bd510a96 & \url{https://github.com/Arachnid/uscc}\\
25     & USCC5 & 3b26643a85d182a9b8f0b6fe8c1153f3bd510a96 & \url{https://github.com/Arachnid/uscc}\\
26     & PW    & 657da22245dcfe0fe1cccc58ee8cd86924d65cdd & \url{https://github.com/paritytech/contracts}\\
27     & BNK   & 97f1c3195bc6f4d8b3393016ecf106b42a2b1d97 & \url{https://github.com/Bankera-token/BNK-ETH-Contract}
\end{tabular}
}
\vspace{1em}
\caption{Smart contract repositories.}
\label{tab:repos}
\end{table*}

\end{document}